\documentclass{emulateapj-rtx4}
\usepackage{amssymb, amsmath, epsfig}

\newcommand{\cmpc}{{\rm cm}^{-3}~{\rm pc}}
\newcommand{\calM}{{\cal M}}
\newcommand{\DM}{{\rm DM}}

\newcommand{\Rvir}{{{\rm r}_{\rm vir}}}

\newcommand{\Msun}{{M_{\odot}}}

\def\apj{ApJ}
\def\apjl{ApJL}
\def\apjs{ApJS}

\def\mnras{MNRAS}
\def\physrep{physrep}

\def\araa{{Ann.\ Rev.\ Astron.\& Astrophys.\ }}

\begin{document}

\title{locating the ``missing'' baryons with extragalactic dispersion measure estimates}
\author{Matthew McQuinn\altaffilmark{1}}

\altaffiltext{1} {Hubble fellow}
      
\begin{abstract}         
Recently, Thornton and coworkers confirmed a class of millisecond radio bursts likely of extragalactic origin that is well-suited for estimating dispersion measures (DMs).    We calculate the probability distribution of DM$(z)$ in different models for how the cosmic baryons are distributed (both analytically and with cosmological simulations).  We show that the distribution of $\DM$ is quite sensitive to whether the ``missing'' baryons lie around the virial radius of $10^{11}-10^{13}~\Msun$ halos or further out, which is not easily constrained with other observational techniques.  The intrinsic contribution to \DM\ from each source could complicate studies of the extragalactic contribution.  This difficulty is avoided by stacking based on the impact parameter to foreground galaxies.  We show that a stacking analysis using a sample of $\sim100$ DM measurements from arcminute-localized, $z\gtrsim 0.5$ sources would place interesting constraints at $0.2-2~$halo virial radii on the baryonic mass profile surrounding different galaxy types.  Conveniently for intergalactic studies, sightlines that intersect intervening galactic disks should be easily identified owing to scattering.  A detectable level of scattering may also result from turbulence in the circumgalactic medium.
\end{abstract}


\section{Introduction}

Approximately $5\%$ of the cosmic baryons at $z\sim0$ are observed to lie within galaxies, $5\%$ are seen as an $X$-ray coronae in massive groups and clusters, and $30\%$ reside in a warm intergalactic phase observed in Ly$\alpha$ absorption  \citep{kauffmann03,fukugita04,penton04}.  The constraints on the locations of the rest (the majority) of the cosmic baryons are weaker, as they reside at densities and temperatures that do not afford significant absorption or emission, except sometimes from highly--ionized states of oxygen \citep{cen99,bregman07,shull11}.

Finding these ``missing'' baryons would inform models for accretion onto, and feedback within, galactic halos.  Half of the Universe's dark matter resides in halos with $m_h>10^{10}\Msun$, where $m_h$ is the halo mass.  However, much less than half of the baryons are observed to lie within these halos:  The $z=0$ stellar mass to halo mass ratio for $m_h=10^{12}\Msun$ has been estimated to be $0.2^{+0.2}_{-0.1}f_b$ \citep{behroozi13}, where $f_b=\Omega_b/\Omega_m$, declining sharply towards both lower and higher masses.  The diffuse galactic gas mass to halo mass ratio is even a few times smaller.  
Much of $f_bm_h$ has been observed as hot intrahalo gas in $m_h\gtrsim10^{14}\,\Msun$ halos \citep{dai10}.  However, for $m_h\lesssim10^{13}\,\Msun$, the bulk of the unseen baryons cannot constitute a hot atmosphere, as it would be thermally unstable \citep{sharma12}.  Most of the baryons associated with these lower mass halos likely lie outside of the virial radius.

Here we consider whether the dispersion measures (DMs) of extragalactic sources could aid this cosmic census.  In contrast to the other observables, every diffuse ionized baryon along a sightline contributes equally to the \DM.  Until recently there was no reason to think that redshifted sources existed for which DM could be measured.   \citet{thorton13} identified a class of out-of-the-plane, highly dispersed (and hence likely extragalactic) millisecond radio bursts, confirming previous indications \citep{lorimer07}.  The four bursts reported in \citet{thorton13} have redshifts of $0.5-1$ if their \DM\ were sourced by the intergalactic medium (IGM).  \citet{lorimer13} forecast that widefield radio interferometers that are presently coming online could detect tens per day of these events. 
In addition, several schemes have been proposed for measuring DM towards time-steady astrophysical sources \citep{lieu13, lovelace13}.  Unfortunately, \citet{hirata13} showed that subtle effects make all such schemes insensitive to DM.

Previous theoretical investigations of the uses of extragalactic \DM\ measurements assumed a homogeneous universe \citep{ioka03,inoue04}.  Here we consider the effects of inhomogeneities, which lead to substantial sightline-to-sightline scatter around the mean $\DM(z)$.  In addition to setting the error bar on measurements of the mean \DM(z), the statistics of this scatter constrain the locations of the ``missing'' baryons.  

The calculations in this Letter assume a flat $\Lambda$CDM cosmological model, consistent with the most recent determinations of \citet{planck13}, and the \citet{sheth02} halo mass function.

\section{sightline-to-sightline scatter in $\DM$}
\label{sec:sigDM}

\begin{figure}
\begin{center}
{\epsfig{file=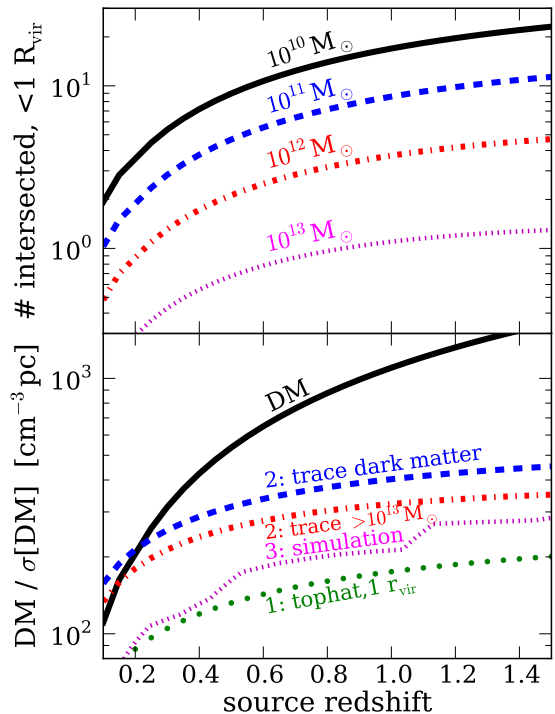, width=8cm}}
\end{center}
\caption{Top panel:  The average number of halos above the specified mass thresholds that a sightline intersects within $1\,\Rvir$.  Bottom panel:  The mean dispersion measure (solid curve) as well as the standard deviation in its value for the considered models (other curves). 
\label{fig:DMvar}}
\end{figure}

Photons propagate through the cosmic plasma at a speed that depends on frequency.  The delay between the arrival time of a photon with observed frequency in units of GHz, $\nu_{\rm GHz}$, and one with a much higher frequency is given by 
\begin{equation}
\Delta t=4.2\;\nu_{\rm GHz}^{-2}\left(\frac{\DM}{10^3\,{\rm cm}^{-3}{\rm pc}}\right){\rm~s},
\end{equation}
where \DM\ is the ``dispersion measure.''  For cosmological distances,
\begin{equation}
{\rm DM}(z_s)=\int_0^{\chi(z_s)}d\chi\,\frac{\rho_e(z,\hat n)}{(1+z)^2} , 
\end{equation}
where $d\chi=c\,dz/H(z)$ is the differential of the conformal distance, $\chi$, $\rho_e(z,\hat n)$ is the electron number density at redshift $z$ in direction $\hat n$, and $z_s$ is the source redshift.  In the bottom panel of Figure~\ref{fig:DMvar}, the solid curve shows the mean value of $\DM(z)$.  For this curve and subsequent analytic calculations, we take \emph{all} of the cosmic baryons to be in a diffuse, fully ionized phase.  

The sightline-to-sightline scatter in $\DM(z)$ primarily owes to scatter in the number of collapsed systems that a sightline encounters.
The top panel of Figure~\ref{fig:DMvar} shows the number of halos above the specified halo mass thresholds that the average sightline intersects within $1\,\Rvir$.  For a sightline with $z_s=1$, on average it intersects $N(m_h)=1$, $3$, $10$, and $20$ halos with $m_h$ greater than $10^{13}$, $10^{12}$, $10^{11}$, and $10^{10}~\Msun$, respectively. The fraction of the dark matter that resides in halos above these masses is $f=0.19$, $0.30$, $0.39$, and $0.46$ at $z=0$ ($f=0.07$, $0.16$, $0.26$, $0.33$ at $z=1$).  
Halos with $m_h<10^{10}\Msun$ are below the Jeans' mass of the IGM and, therefore, unlikely to be overdense in gas.

The sightline-to-sightline variance in $\DM(z_s)$ is given by
\begin{eqnarray}
\sigma^2[{\DM}]&=&\int_0^{z_s}\frac{c\,dz_1}{a_1H(z_1)}\int_0^{z_s}\frac{c\,dz_2}{a_2H(z_2)}\,\bar{\rho}_e^2(0) \left\langle\delta_e(z_1)\delta_e(z_2)\right\rangle,\nonumber\\
&\approx&\int_0^{z_s}\frac{c\,dz}{H(z)}(1+z)^2\bar{\rho}_e^2(0)\int\frac{d^2k_\perp}{(2\pi)^2}P_e(k_\perp,z),\nonumber 
\end{eqnarray}
where $a_n=(1+z_n)^{-1}$, $\bar{\rho}_e(z)$ is the mean electron number density, $\delta_e(z)$ is the electron overdensity, $P_e(k, z)=\langle|\tilde{\delta}_e(k, z)^2|\rangle$ is its spatial 3D power spectrum, tildes denote the Fourier dual in the convention where $2\pi$'s appear only under $dk$'s, and $\langle...\rangle$ indicates an ensemble average.

To calculate $P_e$ and hence $\sigma^2[{\DM}]$, we consider three models (ordered in increasing sophistication) for halos' gas profile of the ionized baryons:  
\begin{enumerate}
\item The baryons associated with $m_h>10^{10}\Msun$ halos are distributed as a top hat with radius $X\,\Rvir$, which yields for each halo a \DM\ at $R\ll X\,\Rvir$ of
\begin{eqnarray}
\Delta_\DM&=&28\,(1+z)\;\frac{\alpha^{2/3}}{X^2}\,\left(\frac{m_h}{10^{12}\Msun}\right)^{1/3}{\rm~cm^{-3}\,pc}.\nonumber
\end{eqnarray}
Here, $\alpha$ is the dark matter density within $1\,\Rvir$ in units of $200$ times its cosmic mean.  The unassociated baryons (or those associated with less massive halos) in this model and Model 2 are assumed to trace the linear density field.  As long as they are more diffuse than the baryons associated with the more massive halos, this assumption has little impact on our results.
\item The baryons trace the dark matter halo profile above a certain mass threshold, $m_*$.  Our calculations assume NFW halo profiles \citep{navarro96} and the concentration--halo mass relation of \citet{bullock01}.  In addition, we use the case $m_*=10^{13}\Msun$ to approximate the \citet{sharma12} model for the intrahalo medium.  \citet{sharma12} find that halos with $m_h>10^{13}\Msun$ have the potential to retain most of their gas in a virialized intrahalo medium, whereas lower mass halos cannot as densities would be required that are thermally unstable.\footnote{We also investigated more sophisticated implementations of the \citet{sharma12} model and found similar results.}
\item The baryon distribution in the ``swinds'' $40\,h^{-1}$Mpc cosmological simulation of \citet{faucher11}, which was run with the GADGET-3 smooth particle hydrodynamics code \citep{springel05}.\footnote{This simulation consists of $512^3$ gas and dark matter particles, with $12$ snapshots between $0<z<1$.  
We estimate that the large-scale modes not captured in this simulation's volume contribute directly $60\,\cmpc$ to $\sigma[\DM]$ for $z_s=1$, which should be added in quadrature to the component from within the volume.}  This simulation uses the \citet{springel03} galactic wind prescription, with $2\,\Msun$ ejected in a $342\,$km\,s$^{-1}$ wind for every $1\,\Msun$ of star formation.  These parameters were chosen to match observations of the $z=0$ stellar mass function.
\end{enumerate}
  
We use {\it the standard halo model} to calculate $P_e$ for Models $1$ and $2$, but with the specified baryonic profiles rather than NFW profiles.  {\it The standard halo model} approximates correlations in the cosmological density field as a superstition of the linear density field correlations (convolved with the halos' profiles) plus a Poissonian term that results from internal correlations within each halo.  This ansatz has met much success reproducing the statistics of the nonlinear dark matter field (see \citealt{cooray02} for a review).  For Model~3, we instead trace skewers through the simulation volume on the light cone, using the simulation's nearest temporal snapshot as the realization of the cosmological density field.

The curves in the bottom panel of Figure~\ref{fig:DMvar} show our estimates for ${\sigma}[{\DM}]$ in the three baryonic profile models.  Model~2 with $m_*=10^{10}\Msun$, annotated as ``trace the dark matter,'' results in the largest dispersion, with ${\sigma}[{\DM}]=400\,\cmpc$ at $z=1$.  The other models have reduced dispersion, with the $1\,\Rvir$ top hat model having the smallest with ${\sigma}[{\DM}]=180\,\cmpc$.  The dispersion in the case where the baryons trace NFW halos for $m_h>10^{13}\Msun$ (which mimics the \citealt{sharma12} model) is only somewhat smaller than the dark matter tracing case, which we explain in Section~\ref{sec:pdf}.  These variances are not only a signal, but set the noise of the stacking analysis discussed in Section~\ref{sec:cross}.  
  
Our models have ignored the contribution of a disky electronic component to ${\sigma}[{\DM}]$.  There are two justifications for this omission.  First, the disky component is unlikely to contribute significantly to ${\sigma}[{\DM}]$.  In the \citet{cordes02} model for the Milky Way electron distribution, an $r=18\,{\rm kpc}$ thick disk contributes a maximum of $60\,\cmpc$ for sightlines perpendicular to the disk plane, and the thick disk is the largest contributor to the electronic column everywhere except in the Galactic Center.  
Consider a toy model motivated by the Milky Way thick disk in which all galactic disks have a column of $\DM_{\rm disk}=100\,\cmpc$.  If $10\%$ of $z_s=1$ sightlines intersect disks (a factor of a few higher than empirical estimates based on damped Ly$\alpha$ systems; \citealt{wolfe05}),\footnote{We can also estimate the probability to intersect a disk analytically. The isothermal potential model in \citet{mo98} yields a disk cross section of $\Sigma_{\rm disk}=\pi(\lambda\Rvir)^2/2\times2/\pi=0.0025~\Rvir^2$, where the latter is evaluated for a halo spin parameter of $\lambda=0.05$ (the RMS found in cosmological simulations).  This model produces a scale radius that is $\sim0.5$ of the estimated termination radius of the Milky Way's electronic thick disk.  The \citet{mo98} model, in conjunction with Figure~1, predicts that a sightline to $z_s=1$ intersects the disk of a $>10^{11}\Msun$ halo $3\%$ of the time.}
  the standard deviation in \DM\ from disks alone would be just ${\sigma_{\rm disk}}[\DM]=30\,\cmpc$, which roughly adds in quadrature to the extragalactic component of $\sim 200\,\cmpc$.   Second, we show in Section~\ref{sec:scattering} that sightlines that intersect interloping disks similar to or denser than the Milky Way thick disk can be distinguished from other sightlines owing to scattering.

\section{probability distribution function of \DM}
\label{sec:pdf}

\begin{figure}
\begin{center}
{\epsfig{file=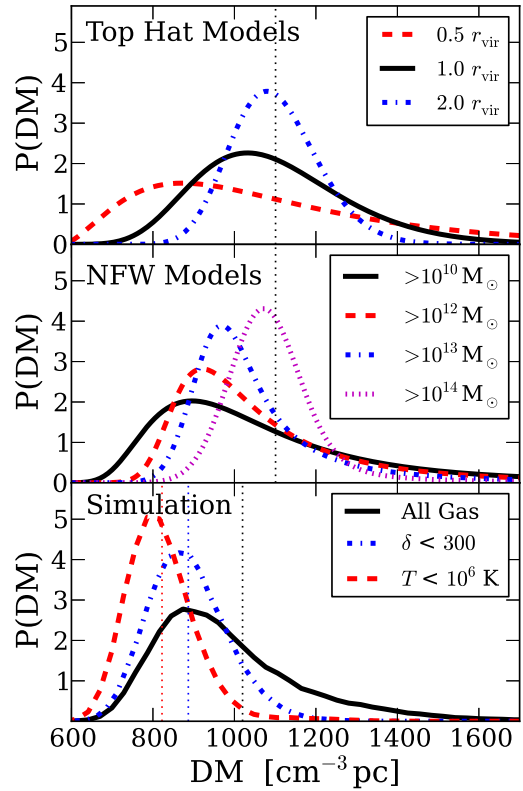, width=8cm}}
\end{center}
\caption{$P(\DM|z_s=1)$ in our analytic models (top two panels) and in the simulation (bottom panel), normalized so that $\int d\DM P(\DM)=1000$.  The different curves illustrate the dependence of this statistic on the extent of the baryonic profile around halos (top panel), on the halo masses that retain their gas (middle), and on baryonic overdensity and temperature thresholds used in the tabulation of \DM\ (bottom).  The dotted vertical lines show the mean \DM.  These lines are slightly offset in the bottom panel owing to star formation in the simulation and/or the specified cuts.
\label{fig:DMpdf}}
\end{figure} 

The characteristic function of the \DM\ distribution -- the Fourier transform of the PDF -- can be calculated in a manner that is similar to the derivation in \citet{zhang07} concerning the PDF of the Compton y-parameter from galaxy clusters.\footnote{This requires the replacement $S\rightarrow\Delta_\DM$ in eqn.~(23) of \citet{zhang07} and integration over the additional parameter $R$.}  For a source at redshift $z_s$, the characteristic function is
\begin{equation}
\widetilde{P}(t |z_s)=\exp\left(\int_0^{\chi(z_s)}d\chi\left[A+\frac{B^2\Delta\chi\,\sigma_{\Delta\chi}^2}{2}\right]\right),
\label{eqn:Pchar}
\end{equation}
where $t$ is the Fourier dual of $\DM$,
\begin{eqnarray}
&A&=\int dm_h\,d^2R\,a^{-2}n(m_h,z)\left(e^{-it\Delta_\DM(R,m_h)}-1\right),\nonumber\\
&B&=\int dm_h\,d^2R\,a^{-2}n(m_h,z)b(m_h,z)\left(e^{-it\Delta_{\DM}(R,m_h)}-1\right),\nonumber\\
&\sigma_{\Delta\chi}^2&=\int\frac{d^2k_\perp dk_\parallel}{(2\pi)^3}P_\delta(k)~{\rm sinc}\left[\frac{\Delta\chi k_\parallel}{2}\right]^2, \nonumber
\end{eqnarray}
$R$ is the proper impact parameter, $\Delta_{\DM}$ is a halo's \DM\ profile at $R$, $n(m_h,z)$ is the comoving number density of halos per $m_h$, $b(m_h,z)$ their linear bias, and $P_\delta$ the linear-theory power spectrum of the dark matter overdensity.
The PDF of ${\DM}$, $P({\DM}|z_s)$, is given by the inverse Fourier transform of $\widetilde{P}(t|z_s)$.  Equation~(\ref{eqn:Pchar}) makes one additional assumption beyond the {\it standard halo model}, that the value of $\delta_e$ for skewers of length $\Delta\chi$ is uncorrelated between adjacent slices.  This is a decent approximation if we evaluate for $\Delta\chi\gtrsim100~$Mpc, which manifests in $\Delta\chi\sigma_{\Delta\chi}^2$ being nearly constant over these $\Delta\chi$. 
  In addition, the diffuse baryonic component that lies far from halos is just the limit of diffuse profiles in equation~(\ref{eqn:Pchar}), $\Delta_\DM(R,m_h)\ll1$, noting that $\int dm\,m/\bar\rho\,n_h(m)=1$.  The form of $\Delta_\DM(R,m_h)$ drops out in this limit.  

 The curves in Figure~\ref{fig:DMpdf} show $P(\DM|1)$ calculated either using equation~(\ref{eqn:Pchar}) or tracing skewers through the simulation.  These panels illustrate its dependence on the extent of the gas profile around halos in the top hat models (top panel), on the specified $m_*$ in the NFW profile models (middle panel), and on the properties of the gas used to tabulate \DM\ in the simulations (bottom panel).  Generically, all the models predict a high--\DM\ tail to the PDF.  This tail is dominated by the most massive systems with $m_h\gtrsim10^{13}\Msun$.  In addition, the more diffuse the gas around halos or the rarer the halos that can hold onto their gas, the more concentrated is $P(\DM|z_s)$.  This trend simply owes to each sightline intersecting a more statistically representative set of structures in the models where the halos' baryonic profiles are more diffuse.  Lastly, the core of $P(\DM|1)$ asymptotes to a Gaussian with $\sigma\approx100\,\cmpc$ that is determined by large-scale cosmological density correlations in the limit that the baryonic profiles are very diffuse (e.g., $2\,\Rvir$ case, top panel) or that most sightlines do not intersect a gaseous halo (e.g., $>10^{14}\Msun$ case, middle panel). 
 
 The simulation allows us to explore how gas at different temperatures and densities contributes to the shape of $P(\DM|z_s)$.  The dash-dotted curve in the bottom panel of Figure~\ref{fig:DMpdf} excludes gas with $\delta_e>300$ from the tally of \DM.  The dashed curve excludes gas with $T>10^6K$.  The comparison of these curves with the solid curve, which includes all of the gas, shows that hot, dense gas (likely associated with $>10^{12}\Msun$ halos) contributes to the high--\DM\ tail in the simulations.  We have also examined the simulation in \citet{faucher11} with winds turned off:  The high--\DM\ tail essentially disappears (and $\sigma^2[{\DM}]$ is halved) in this simulation probably because of overly-efficient star formation.

\section{The DM--galaxy cross correlation}
\label{sec:cross}

\begin{figure}
\begin{center}
{\epsfig{file=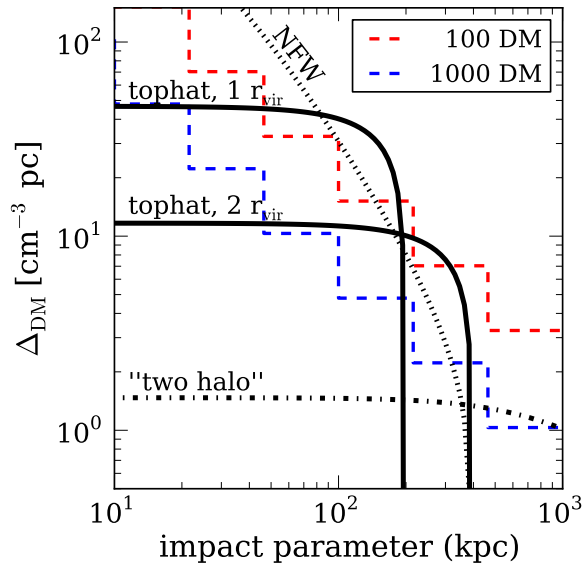, width=8cm}}
\end{center}
\caption{Sensitivity to the average \DM\ profile around $m_h\sim10^{12}~\Msun$ halos of an analysis that stacks sightlines based on their proximity to neighboring galaxies.  The solid curves are the \DM\ profile of a $z=0.5$, $10^{12}~\Msun$ halo if the baryons were distributed as a top hat with radius $1$ and $2\,\Rvir$.  The dotted curve uses the same specifications except the baryons trace the dark matter with an NFW profile truncated at $2~\Rvir$.  The dashed piecewise lines are the expected $1\sigma$ errors for a survey with $10^2$ and $10^3$ \DM\ measurements with $z_s\sim0.5-1$, assuming $\overline{\sigma_{\DM}}=200\,\cmpc$ and $\ll1'$ localizations.  
\label{fig:DM_profile}}
\end{figure}

It is not possible to separate the \DM\ contribution that is intrinsic to the sources from that which is extragalactic.  However, the intrinsic contribution does not contaminate the signal if sightlines are stacked based on their impact parameters to foreground galaxies.  In addition, stacking does not require precise knowledge of the sources' redshifts.  Upcoming photometric surveys such as DES and LSST aim to identify most $z\lesssim1.5$ galaxies over $\sim0.5$ of the sky.\footnote{\url{http://www.darkenergysurvey.org}, \url{http://www.lsst.org/}}  Targeted follow-up along each \DM\ sightline is also a possibility.  Stacking requires a resolution of $<0.5'$ (or $\gtrsim1\,$km baselines at $1~$GHz) to localize the source to $<1\,\Rvir$ of a foreground $z=0.5$, $10^{12}\,\Msun$ halo. 

The enhancement in \DM\ from a local galaxy with halo mass $m_h$ at impact parameter $R$ is
\begin{equation}
{\Delta_{\DM}}=\Delta_{\rm1h}(R,m_h)+2\,\int_{r_{\rm min}}^\infty\frac{d\Delta\chi}{a}\, \zeta_{2h}(\sqrt{\Delta\chi^2+(R/a)^2},z),\nonumber
\end{equation}
where $\Delta_{\rm1h}(R,m_h)$ is the dispersion measure profile of the halo, $\zeta_{2h}=\rho_e(0)b_h(m_h)\zeta_\delta$, and $\zeta_\delta$ the 3D matter overdensity correlation function.  We calculate the latter, ``two halo'' term with $r_{\rm min}/\Rvir=1.5$ and assuming linear theory as in the {\it standard halo model} (but highly approximate at such small $r_{\rm min}$).

In a survey with $n_{\DM}$ sightlines, the maximum likelihood $\Delta_{\DM}(R,m_h)$ estimator for a Gaussian $P(\DM)$ is
\begin{equation}
\widehat{\Delta}_{\DM}=\sum_i^{n_{\DM}}\frac{N_i}{\sigma_{\DM,i}^2} \left(\widehat{\DM}_i-\overline{\widehat{\DM}} -\delta(z_i) \right)/\sum_i^{n_{\DM}}\frac{N_i}{\sigma_{\DM,i}^2}\left(N_i-\overline{N_i}\right),\nonumber
\end{equation}
where $\overline{Y}=(\sum_i\sigma_{\DM,i}^{-2})^{-1}\sum_iY_i\times\sigma_{\DM,i}^{-2}$ for any variable $Y_i$, $\widehat{\DM}_i$ is the estimated \DM\ to source $i$, $N_i$ is the number of halos along the sightline that fall in the $(R,m_h)$ bin, $\delta(z_i)\equiv\langle\DM\rangle(z_i) -\overline{\langle\DM\rangle}$, and $\sigma_{\DM, i}^2$ is the variance around the mean $\DM(z_i)$.  This estimator assumes that $\delta(z)$ is known, but can easily be generalized to fit a parametric form.  The standard deviation of this estimator is
\begin{equation}
{\rm\sigma}[\widehat{\Delta}_{\DM}(R,m_h)]=\sqrt{\frac{\overline{\sigma^2_{\DM}}}{\left(\overline{N^2}-\overline{N}^2\right)\,n_{\DM}}}\approx \sqrt{\frac{\overline{\sigma_{\DM}^2}}{\overline{N}\,n_{\DM}}}.\nonumber
\end{equation}
This error would be reduced if other $\widehat{\Delta}_{\DM}(R,m_h)$ bins are estimated simultaneously.

Figure \ref{fig:DM_profile} shows the sensitivity of such a stacking analysis to the mean \DM\ profile of $10^{12}~\Msun$ halos.  The dashed piecewise lines are the $1\,\sigma$ error for a survey with $n_{\DM}=10^2$ and $10^3$ in a radial bin given by the width of each horizontal line segment.  This calculation assumed $\overline{\sigma_{\DM}}=200\,\cmpc$ and that each sightline on average intersects two $10^{12}~\Msun$ halos within $1\,\Rvir$.  These numbers roughly approximate our Model~3 for $z_s=0.5-1$ if the intrinsic dispersion in \DM\ is somewhat smaller than the cosmological dispersion (see Figure~\ref{fig:DMvar}).  
Figure~\ref{fig:DM_profile} shows that both the $n_{\DM}=10^2$ and $10^3$ cases are sensitive to the $1\,\Rvir$ top hat model -- a model that cannot be ruled out with other techniques \citep{fang13} -- and the survey with $n_{\DM}=10^3$ is even sensitive to the $2\,\Rvir$ model.  The sensitivity to ${\Delta}_{\DM}(R)$ improves [degrades] for a stacking analysis around less [more] massive halos compared to our fiducial $10^{12}\Msun$ case owing to the different $\overline{N}$.  However, this trend is offset (almost perfectly in the top hat models) by the likely increase in ${\Delta}_{\DM}$ with halo mass.

\section{scattering}
\label{sec:scattering}

Scattering (which broadens an electromagnetic pulse in a frequency--dependent manner owing to path length variations) could result in \DM--dependent selection effects.  
In fact, any millisecond pulse whose sightline intersects an intervening galactic disk  is likely to be significantly broadened, over a timescale of
\begin{eqnarray}
\tau_{\rm sc}^{\rm disk}&=&90{\rm~ms}\times{\rm SM}_{-4}^{6/5}\,\nu_{\rm GHz}^{-22/5}\label{eqn:taudisk}\\
&&\times\left[\frac{\chi(z)}{1{\rm\,Gpc}}\right]\left[1-\frac{\chi(z)}{\chi(z_s)}\right]\left(1+z\right)^{-22/5},\nonumber
\end{eqnarray}
where ${\rm SM}_{-4}$ is the ``scattering measure'', in units of $10^{-4}{\rm m^{-20/3}kpc}$ \citep{blandford85,cordes02}.  Equation~(\ref{eqn:taudisk}) is calculated using that $2c\tau_{\rm sc}=\chi\theta(1-\chi/\chi_s)$ to a cosmological source, where $\theta$ is the RMS deflection angle owing to scattering (see equation~4.23 in \citealt{blandford85}).  
Equation~(\ref{eqn:taudisk}) assumed the density power spectrum of Kolmogorov turbulence, as does what follows.  For pulsar sightlines above the Milky Way disk plane, ${\rm SM}_{-4}\sim1$, and these sightlines are characteristic of the Milky Way's thick electron disk \citep{cordes02}.  Thus, $z\lesssim1$ millisecond transients whose sightline intersects a gaseous disk similar to the Milky Way are significantly broadened at $\nu_{\rm GHz}\lesssim1$.

Sightlines passing through a halo's outskirts are likely to be less broadened by scattering with
\begin{eqnarray}
{\tau}_{\rm sc}^{\rm halo}&=&5~{\rm\mu s}\times\nu_{{\rm GHz}}^{-22/5}\left[\frac{1{\rm\,kpc}}{\ell_{\rm out}}\right]^{4/5}\left(\frac{\calM\rho_{e,\rm char}}{10^{-5}{\rm\,cm}^{-3}}\right)^{12/5}\label{eqn:tauhalo}\\ 
&&\times\left(\frac{L_{\rm halo}}{10^2{\rm\,kpc}}\right)^{6/5}\left[\frac{\chi(z)}{1{\rm\,Gpc}}\right]\left[1-\frac{\chi(z)}{\chi(z_s)}\right]\left(1+z\right)^{-22/5},\nonumber
\end{eqnarray}
where $\ell_{\rm out}$ is the maximum turbulent eddy scale (determined by the driving mechanisms; e.g., infall, winds) and $\calM\,\bar\rho_e$ is the standard deviation in the electron number density.  For mildly subsonic turbulence, $\rho_{e,\rm char}$ can roughly be thought of as the characteristic density and $\calM$ the typical compressive Mach number \citep{hopkins13}.  Simulations find $\calM\sim0.1-0.2$ at $<2\Rvir$ for halos with $m_h\sim10^{15}\Msun$ \citep{lau09}, and it is likely that similar Mach numbers apply at lower $m_h$.  For our fiducial parameters -- the parameter choices that yield the $5~{\rm\mu s}$ coefficient in equation~(\ref{eqn:tauhalo}) -- ${\tau}_{\rm sc}^{\rm halo}$ is larger than the Milky Way contribution of ${\tau}_{\rm sc}\sim0.1\,\nu_{\rm GHz}^{-22/5}\mu{\rm s}$ for high-latitude sources \citep{cordes02}.  Moreover, ${\tau}_{\rm sc}^{\rm halo}$ is likely measurable for the \citealt{thorton13} bursts:  Setting $\nu=250$MHz in equation~(\ref{eqn:tauhalo}) results in $5\mu{\rm s}\rightarrow2$ms.

Equation~(\ref{eqn:taudisk}) and (\ref{eqn:tauhalo}) assumed strong scattering (a requirement for multi-path propagation from a point-source), defined as when the Fresnel scale is larger than the transverse separation at which a ray's RMS phase differs by $\pi$, $r_{\rm diff}$.  In the Milky Way ISM, the ratio of these scales is $f\sim10^2\nu_{\rm GHz}^{-17/10}$ such that strong scattering applies.  For the fiducial intrahalo parameters in equation~(\ref{eqn:tauhalo}), this condition is similarly satisfied with 
\begin{equation}
f=1\times 10^{2}\nu_{\rm GHz}^{-17/10}a^{11/5}\left(\ell_{\rm out}^{-2/5}[\calM\rho_{e,\rm char}]^{6/5}L_{\rm halo}^{3/5}\chi^{1/2}\right)_{\rm fiducial}.\nonumber
\end{equation}

Equation~(\ref{eqn:tauhalo}) also assumed that the inner scale of turbulence satisfies $\ell_{\rm in}<r_{\rm diff}=2\times10^{12}\nu_{\rm GHz}^{6/5}a^{-6/5}$~cm, much larger than the limit $\ell_{\rm in}<10^{10}\,$cm found in the Milky Way ISM \citep{armstrong95}.  However, this condition may not be satisfied in the circumgalactic medium.  \citet{lithwick01} argued that $\ell_{\rm in}$ is set by the eddy scale at which the turnover time equals the proton diffusion time across the eddy or
\begin{equation}
\ell_{\rm in}\sim6\times10^{13}\left(\frac{1{\rm\,kpc}}{\ell_{\rm out}}\right)^{1/2}\left(\frac{10^{-4}{\rm\,cm}^{-3}}{\rho_{e}/\beta}\right)^{3/2}{\rm~cm},
\end{equation}
where $\beta$ is the plasma beta parameter (the ratio of thermal to magnetic pressure), which is likely $>1$ in the intrahalo medium.  However, it is also possible that the inner scale is set by the much smaller proton-gyro radius \citep{schekochihin09}.\footnote{In $\beta\gg1$ Alfv\'enic turbulence it is likely that density fluctuations are suppressed relative to velocity fluctuations by $\beta$, resulting in smaller $\tau_{\rm sc}^{\rm halo}$ than predicted by equation~(\ref{eqn:tauhalo}).}
  If $r_{\rm diff}<\ell_{\rm in}$, $\tau_{\rm sc}^{\rm halo}$ would be reduced by the factor $\sim(r_{\rm diff}/\ell_{\rm in})^{1/3}$ as long as the scattering remained strong.

Pulse broadening with the expected frequency dependence for scattering was detected in two fast radio bursts: one of the four reported in \citet{thorton13}, with $\widehat{\tau}_{\rm sc}=3.7\,$ms  at $1$GHz \citep{lorimer13}, and the \citet{lorimer07} burst with $\widehat{\tau}_{\rm sc}=20\,$ms.   These $\widehat{\tau}_{\rm sc}$ suggest that either (1) the scattering is intrinsic or (2) the sightline intersected a galactic disk.  However, the latter seems unlikely unless electronic disks have significantly larger radii than we have estimated (see footnote 4).

\section{Conclusions}

 We showed that the standard deviation around the mean $\DM(z)$ is $100-400\,\cmpc$ at $z=0.5-1$, with its exact value being sensitive to how the baryons are distributed.  
Extragalactic \DM\ measurements could constrain the locations of the cosmic baryons either by directly measuring the probability distribution of the intergalactic \DM\ or by stacking based on the separation to field galaxies.  The former method requires both the sources' redshifts as well as knowledge of their intrinsic $\sigma[\DM]$ to a precision of $\ll400\,\cmpc$, whereas the latter requires sub-arcminute localization and the identification of coincident galaxies.\\
 
We thank Chris Hirata, Ryan O'Leary, Eliot Quataert, Martin White, and Jennifer Yeh for helpful conversations, and Claude-Andr{\'e} Faucher-Gigu{\`e}re for sharing the simulations used in this work.  MM acknowledges support by the National Aeronautics and Space Administration through the Hubble Postdoctoral Fellowship.

\bibliographystyle{apj}

\end{document}